# The neuron as a temporal electroacoustic medium.

F. Meseguer, and F. Ramiro-Manzano

*Instituto de Tecnología Química (CSIC - UPV),*
*Universitat Politècnica de València, Av.Tarongers s/n 46022, Valencia, Spain*

**The human brain is one of the most complex and intriguing scientific topics. The most established theory on neuronal communication is a pure electrical model based on the propagation of intracell cationic charges along the neurons. Here we propose a complementary model based on two properties of brain communication: A) The Coulomb interaction associated to the Action Potential (AP) pulse induces a deformation of the neuron membrane which travels as an acoustic signal, i.e.: The ions play an essential role and the electric and acoustic signals, composing the AP, are strongly correlated. B) As brain communication is stablished through a periodic train of AP pulses it induces a time periodic modulation of the acoustic parameters. In this framework we propose envisaging the neuron as a temporal electro-acoustic medium. The temporal varying media framework could help understanding brain conundrums such as propagation routes involved in the neuronal plasticity in the consolidation of the memory, as well as on the generation of the signals associated to the brain field theory.**

The brain is the most complex system of the human body. Brain activity is distributed over an intricate network of neurons (approximately $10^{11}$), where the transmission and management of information is carried out through neural connections (about $10^3$ per neuron) called synapses. One of the most important subjects of inquiry in neurology concerns understanding how the brain can receive, manage, and record experiences and sensations, as well as later evoke memories and emotions. The extraordinary process of integration is performed through the communication between different areas of the brain in a gigantic network some authors called it as the brain web [1] or the connectome [2,3].

The more accepted model for brain communication is a pure electrical model developed by Hodgkin and Huxley (HH) in 1952 [4]. However, first in 1980 Tasaki and collaborators [5] and later other groups around the world [6,7,8,9], reported on the swelling of nerve fibers associated to the Action Potential (AP) pulse, pointing out to an acoustic wave propagating together with the electrical pulse. From a schematic vision, the propagation of the acoustic signal in a neuron would be like the sound propagation along a cylindric shaped long membrane filled with the neuron fluid (essentially an ionic solution) immersed in practically the same neuron fluid. Due to the different acoustic parameters of the wave components of the AP pulse the pulse propagation is usually distorted and at a certain distance its integrity is compromised. In 2009 Heimburg and Jackson [10] proposed an acoustic wave model based on soliton transmission, addressing this problem. Here, the non-lineal properties of the medium, attributed to a phase change of the lipidic membrane, compensate the dispersion. In this way the shape of the pulse is preserved and the soliton is propagated. Later, Schneider et al. provided experimental support to solitary solutions [11,12,13]. Although the soliton model is a very interesting proposal it does not include the role of the intracell ion cloud. Recently other models based on surface/membrane waves, connect with the HH model though interchange of electrical potential and kinetical energy [ 14 ] or flexoelectricity [15].

Here we present an electroacoustic model of neuronal communication based on the physicochemical properties associated to the action potential (AP) pulse. Although this model is naïve and it must be experimentally tested, we believe it presents all ingredients to be considered as a framework of discussion which might provide some hints for unveiling the complex nature of brain processes. The model is based on two closely related key points of brain communication: A) The AP induces a Coulomb interaction in both the intracell and extracell $Na^+$ ions near the neuron membrane. It triggers a mechanical deformation of the neuron membrane. However, and more important, Coulomb interaction within de ion cloud induces a change of the fluid stiffness responsible of the acoustic wave propagation, i.e.: the electrostatic interaction produces a temporal modulation of the acoustic parameters associated to the AP pulse. This model might also give arguments for understanding the heating-cooling cycle appearing in the AP pulse[6]. B) Brain communication is stablished through a train of pulses. Assuming that, (i) the medium is perturbed and modulated by these acoustic pulses and considering that (ii) these signals are periodic or quasiperiodic in time, the neuron could be envisaged as a temporal electro-acoustic medium. In the literature, time varying medium has been postulated as a framework for obtaining a plethora of wave phenomena such as dynamic changes of wave propagation [16], propagation through the border (edge states) [17], wave generation [18], resonant behavior [19] and amplification [19,20,21]. Our aim concerns showing the neuron might be a temporal electro-acoustic medium candidate which may help understanding the complex phenomena of brain communication.

**Results and discussion**

**1. Electro-acoustic coupling.**

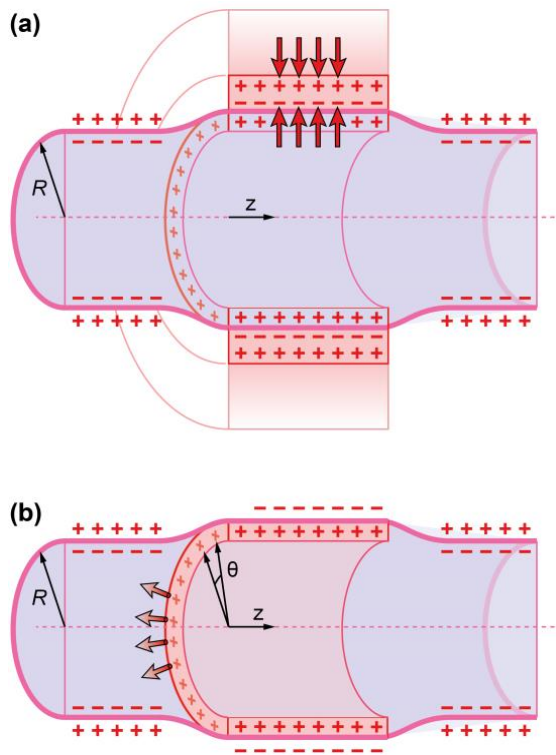

*Figure 1. Two scenario models of Na+ charge distribution in un-myelinated neurons: a) Debye layer extracell Na+ ion cloud; b) intracell Na+ ion cloud.*

Most models are based on the seminal work of Hodgkin and Huxley which assume the extracellular fluid is electroneutral and the only contributions to the electric field is given by membrane currents [22]. However other models call attention to the role of $Na^+$ cations located at extracell Debye layer [23].

Under the AP pulse, a measurable deformation of the neuron membrane has been reported [5, 24, 25]. Here, we postulate the origin of this mechanical wave is based on the Coulomb interaction forces of intracell and/or extracell ions. However, as the biophysical basis of the AP signals is still not fully understood, we propose two possible complementary ion distribution scenarios both of them contributing to the deformation of the neuron membrane (see Figure 1).
A) Neuron deformation is originated from the extracellular Debye layer of $Na^+$ ions attached at the outer side of the neuron membrane [23].
B) Neuron deformation is originated from the electrostatic repulsion [26] within the intracellular layer of $Na^+$ ions at the inner side of the neuron membrane.
Considering that the ionic layer thickness of the model B is similar to Debye layer of the model A it might result on a maximum ion density value around $\zeta \approx 6.3 \cdot 10^{18}/cm^{-3}$. It produces a time dependent ion concentration pulse this propagating along the neuron. To understand how large this figure is, let us compare it to the density of electrons appearing in well-known materials used in Electronics and Material Science [26]. The [$Na^+$] ion concentration within the action potential pulse is ten times larger than the impurity concentration of a doped degenerate semiconductor like silicon [26]. In both models the $Na^+$ ions shell push out the neuron membrane. Consequently, it induces a transversal acoustic wave along the neuron. Also, as the neuron is immersed in the extracellular brain fluid both intra and extracellular $Na^+$ layers induce a stiffening effect on the neuron fluid. Therefore, they should be considered in the calculation of the stiffness value, a key parameter which controls the velocity of the AP acoustic wave. In the scenario A (Figure 1a) the un-myelinated neuron model is built up by a cylindrical membrane filled with the neuron fluid, (essentially water) immersed in the extracellular fluid, essentially water plus a Na+ cloud located at the Debye layer. However in the scenario B the Na$^+$ cloud is located in the intracellular fluid. As fluid presents zero shear elasticity, we assume the relevant factors triggering the transversal acoustic wave propagation are both, the compressibility parameters of the Na$^+$ ion layer, $\kappa_{ion}$, and the surface compressibility of the lipid membrane $\kappa_A$.

Therefore, the total compressibility, $\kappa$, of the neuron can be written as,

$$1/\kappa = 1/\kappa_M + 1/\kappa_{ion} \quad (1)$$

Where $\kappa_M$ and $\kappa_{ion}$ are the compressibility parameters of the neuron membrane and the Coulomb induced compressibility of the Na ion cloud respectively. As both scenarios A, and B contribute to the compressibility parameter $\kappa_{ion}$, then,

$$1/\kappa_{ion} = 1/\kappa_{ion}^A + 1/\kappa_{ion}^B \quad (2)$$

The sound velocity, *c*, in a fluid is then given by the following expression [27]:

$$c = 1/\sqrt{\kappa\rho} \quad (3)$$

being $\rho$, the density value of the fluid medium. Table 1 shows the compressibility values $\kappa^A_{ion}$ and $\kappa^B_{ion}$ for both models A and B as well as the total compressibility $\kappa_{ion}$. It also shows the acoustic velocity at the maximum value of the AP (full depolarization) as well as half-depolarization obtained through the expressions developed in the Supporting Information. From Table 1 we see $\kappa_{ion}^B \gg \kappa_{ion}^A$. Also as the compressibility value of the neuron surface membrane takes very large values [10,28], ($\kappa_M(37^\circ C) = 5{,}4 \, (Pa)^{-1}$ we conclude that $\kappa_M \gg \kappa_{ion}^B \gg \kappa_{ion}^A$. Consequently $\kappa \approx \kappa_{ion}^A$, and $c \cong 1/\sqrt{\kappa_{ion}^A \rho}$.

| Cell membrane | Q (C/cm$^2$) | *d* (nm) | $\kappa^A_{ion}$ (Pa)$^{-1}$ | $\kappa^B_{ion}$ (Pa)$^{-1}$ | $\kappa_{ion}$ (Pa)$^{-1}$ | c (m/s) |
|---|---|---|---|---|---|---|
| depolarized | 1.0·10$^{-7}$ | 12.6 | 2.1·10$^{-5}$ | 1.1·10$^{-3}$ | 2.0·10$^{-5}$ | 7.0 |
| half-depolarized | 5.0·10$^{-8}$ | 17.9 | 4.2·10$^{-5}$ | 4.3·10$^{-3}$ | 4.1·10$^{-5}$ | 4.9 |
| polarized | 1.0·10$^{-8}$ | 40.0 | 2.1·10$^{-4}$ | 1.1·10$^{-1}$ | 2.1·10$^{-4}$ | 2.2 |

*Table 1. Fundamental parameters evaluated for calculating the sound speed. Q (Coulomb/cm$^2$) is the Na$^+$ ion concentration and d represents the nearest neighbor Na$^+$ ion distance in nanometers. It also shows the compressibility parameters of both models, as well as the total compressibility $\kappa_{ion}$ and the phase velocity, c.*

We have assumed a depolarization voltage value of 0.1 Volts and a membrane capacitance, C = 1.0·µF/cm$^2$. It is important emphasizing those models are very simple since they do not consider other properties of the neuron membrane as its surface viscosity. It is important pointing out that, in both scenarios $\kappa_{ion}$ is not a constant parameter, but it is a time dependent parameter which value is assumed to change about one order of magnitude during the AP pulse (see point 2, time varying electroacoustic medium). As the acoustic properties of the neuron varies in time we consider that the mean value of phase velocity corresponds to the half-depolarization of the AP pulse (see Table 1); i.e.: A phase velocity of the AP pulse, *c* = 4.9 m/s, of the same order of magnitude of those reported for un-myelinated neurons [29].

This model might also give response to the large velocity values of the AP pulse in the case of myelinated neurons. Myelin covering layer produces a mechanical hardening of the neuron surface this probably decreasing the surface compressibility factor $\kappa$, but not as small as the compressibility due to the Coulomb interaction of the ion cloud. The concentration of Na$^+$ channels, *N*, at the nodes of Ranvier increases enormously the intracell Na$^+$ ion flux producing a dramatic increase of the neuron deformation. It is known that the density of Na$^+$ channels in myelinated axons, at the nodes of Ranvier, is around 26 times larger than those of unmyelinated axons [30,31,32]. If we assume all Na$^+$ ion channels provide the same ion flux, the ratio between the pulse velocity of myelinated and unmyelinated axons would be (see Supp. Information)

$$c_{myel}/c_{unmyel} \cong c_{myel}^A/c_{unmyel}^A = \left(N_{myel}/N_{unmyel}\right)^{1/2} = 5.1 \quad (4)$$

i.e.: the AP pulse in myelinated axons should travel around 5 times faster than that of unmyelinated axons. Table 2 shows the calculated values of the compressibility parameter $\kappa$ and the mean values of the phase velocity for unmyelinated and myelinated neurons. For more details on the calculation of the stiffness coefficient see the Supporting Information.

| Axon | $\kappa$ (Pa)$^{-1}$ | c(ms$^{-1}$) |
|---|---|---|
| Unmyelinated (half-depolarized) | $4.1 \cdot 10^{-5}$ | 4.9 |
| Myelinated (half-depolarized) | $1.6 \cdot 10^{-6}$ | 25.0 |

*Table 2. Compressibility parameter κ, see equation (S3), and phase velocity, c, obtained from equation (3)*

We can see the large differences in the value of the compressibility parameter $\kappa$ for myelinated and unmyelinated axons at the depolarized stages of the AP. We must emphasize these are very simple approaches which gives a rough estimation of the pulse velocities of the same order of magnitude as those experimentally observed [29].

The electroacoustic model might also help understanding the so call "saltatory conduction" in myelinated neurons [33]. To the best of our knowledge, conduction models so far published are based on the decrease of the transverse resistance and, more importantly the reduction of the transverse capacitance [34]; i.e.: they are based on cable theory.

However our model predicts a deformation of the neuron membrane which due to flexoelectricity[13] induces a change of the membrane potential. Therefore, the acoustic wave component of the AP can couple neighbor Ranvier nodes, enabling the AP propagation; i.e.: the saltatory conduction can be envisaged as a tunneling effect between Ranvier nodes, this enabling the AP propagation. A very simple calculation shows that the acoustic wavelength $\lambda$ of the carrier wave of the AP can roughly be estimated by the following expression

$$\lambda = c\Delta T \quad (5)$$

Where $\Delta T$ is the AP time amplitude (of the order of $\Delta T = 0.5\ ms$). If we assume a pulse velocity value of c = 25 m/s we get a wavelength value of $\lambda \approx 1\ cm$. In the framework of tunneling effect, $\lambda$ defines the exponential decay, which is much larger than the internodal distance of myelinated neurons(of the order of 1mm). Therefore, the acoustic component of the AP pulse might trigger opening very fast Na$^+$ channels of the neighbor forthcoming nodes of Ranvier.

The model can also explain the origin of the heating/cooling cycle associated to the AP signal [6]. Several interpretations to the observed heating/cooling cycle have been published [6,35,36]. Of especial interest is the proposal given by Abbott et al. in 1958 [35], and later revisited by Chandler et al. [37]. This model which was able to partially account for the heat production/absorption is based on the free energy charge/discharge cycle of the neuron membrane capacitor. Our model goes on the line of the Abbott proposal, but it suggests including the compression/expansion cycle of the Na$^+$ ion cloud as an additional mechanism for the heat emission/absorption mechanism; i.e.: the intracell Na$^+$ ion accumulation produced by the AP pulse can be envisaged as a gas cloud submitted to a compression/expansion process which might result on a heating/cooling Carnot cycle like it appears in the refrigerant gas of a fridge.

**2. Time varying electroacoustic medium.**

Brain communication is stablished through a train of AP pulses which customarily are periodic or quasiperiodic in time. As the AP pulse strongly modifies the compressibility parameter of the intracell fluid, the pulse train produces a quasiperiodic modulation of the acoustic properties of the neuronal fluid. So, the sound velocity $c$ in the medium and the compressibility $\kappa$ parameters are time periodic functions i.e.:

$$c(t) = c(t+T), \kappa(t) = \kappa(t+T), \quad (4)$$

being $T$ the firing rate of the AP pulse.

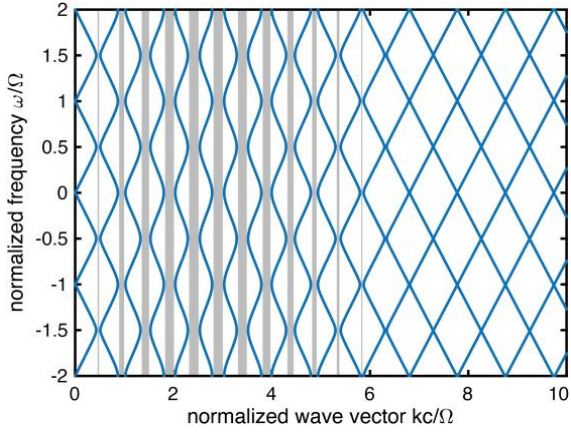

Figure 2. Frequency ω vs k-vector dispersion relation in normalized units for the relative compressibility parameter $\kappa'_M = 0.10$

The propagation of waves in media whose parameters are periodic in time has been of great interest for the Metamaterials community. They are called as Temporal Photonic Crystals [18,19, 38, 39, 40] or Photonic Time Crystals [20,21] in the case of Electromagnetic Waves, and Temporal Phononic Crystals [41,42] in the case of acoustic waves. However, this research topic was for many years of great interest to the electronic engineering community under the name of Time Varying Dielectric Media [43,44,45].

One of the fingerprints of a time varying media is the appearance of certain wave-vector regions, called as gaps where the wave is not allowed to be propagated through it. These gaps also appear in media whose either optical or acoustic properties are periodic in the space [46,47], and they are called as Photonic Crystals [48,49] or Sonic Crystals [50] respectively. However here gaps appear in the frequency region.

As a neuron can be modelled as a narrow and long cylinder of radius R (see Figure 1), we assume the acoustic signal is propagating along a one-dimensional (1D) system. Here, the membrane swelling is produced by the sound pressure $p_R(z,t)$ across the radius and it propagates along the cylindrical axis of the neuron (z direction).

Then, by analogy to the Electromagnetic (EM) case the sound wave equation can be written as [18,27];

$$\frac{\partial^2}{\partial z^2} p_R(z,t) - \frac{\partial^2}{\partial t^2}\left[p_R(z,t) \big/ \big(c(z,t)\big)^2\right] = 0 \quad (5)$$

$$c^2 = 1/\rho\kappa = c_0^2 \kappa_0 / \kappa \quad (6)$$

$c_0$ and $\kappa_0$ being the sound velocity and the compressibility of the intracellular fluid at the polarized stage, respectively. Here the fluid density, $\rho$, is a constant parameter (we can roughly assume the intracell fluid is essentially water).

After an algebraic development like that of Temporal Photonic Crystals [19] (see de Supp. Information) we obtain the following canonical equation:

$$\sum_m \left[\sum_{p-m} \alpha_p(\omega) \kappa'_{p-m}(\omega) - k^2 c_0^2 \delta_{pm}\right] q_m(\omega) = 0 \quad (7)$$

$$p, m = 0, \pm 1, \pm 2, \pm 3 \ldots$$

$$\alpha_p(\omega) = (\omega - p\Omega)^2$$

being $\delta_{pm}$, $\kappa'_{p-m}(\omega)$ and $q_m(\omega)$ the Kronecker delta tensor, the Fourier components of the relative compressibility, and the sound wave pressure amplitude respectively. Here we have assumed that $\kappa' = 0.10$ (see Fig S.1a).

Equation (7) can be regarded as the interaction of a transverse acoustic signal of frequency $\omega$ propagating in the intracell medium (the axon) with a compressibility, $\kappa'(t)$, periodic in time originated by the periodic train of pulses of frequency $\Omega = 2\pi/T$.

After introducing the Fourier components of the relative compressibility parameter (see the Supp. Information) into equation (7), we get the frequency vs $k$-vector value of the acoustic modes in the neuron. Figure 2 shows the solution of equation (7); i.e.: it shows the dispersion relation $\omega(k)$ of the propagating neuronal acoustic modes (figure 2). This result is like the dispersion obtained in a transmission line of Reyes-Ayona and Halevi [51], but with some specific properties different to that of Photonic Time Crystal. In the following paragraph we explain and discuss the obtained results.

The blue lines of Figure 2 represent the available acoustic wave parameters ($\omega$ vs $k$-wavevector) or solutions of the acoustic wave travelling through the neuron. These solutions show the typical periodic pattern in the vertical axes (frequency). In the literature where usually, the medium is modulated in time by a sinusoidal wave, the period of this pattern corresponds to its single frequency. By contrast, the modulation in this study is created by pulses, each one composed by multiple waves (see Fig S2b). In contrast to the usual single gap of the literature, here multiple gaps (gray regions) have been obtained. This is due to both, the presence of multiple wave components of the AP pulse (Fig. S2b) as well as the strong modulation of the compressibility parameter (Fig. S2a).

Next, we will discuss some physiological consequences of considering the neuron as time-varying medium. The presence of the time-varying electroacoustic medium represents a dynamic and malleable behavior of the media. In particular, the peculiarities of the train of pulses would modify or determine the wave propagation, including effects of reflection suppression [52] and/or amplification [19]. i.e.: the temporal crystal configuration could dynamically influence different responses at axon branch points, such as reflection, conduction block and full conduction [53,54], as well as amplification effects [19,55]. It might produce cellular modifications as was first visioned by Donald Hebb [56] and later demonstrated by Eric Kandel [57] and many others concerning neural networks in the process of the consolidation of the memory [58,59,60,61].

In line with the brain field theory, the multidirectional combined response of time varying electro-acoustic media could give rise to collective effects. Consequently, frequency generation, long range resonances and amplification [18,19] could be revealed. This might give theoretical support to the resonant modes appearing in the brain field theory for mapping brain activity [62,63].

## 2. Conclusions

In conclusion, here we present an electroacoustic model of neuronal communication based on two properties of the AP pulse.

A) The Coulomb interaction of the intracell/extracell $Na^+$ ion cloud at the AP pulse induces a mechanical deformation of the neuron membrane, this triggering an acoustic wave which propagates together with an electric pulse, i.e.: both the electric and the acoustic signal are strongly correlated.

B) The periodic distribution of the train of pulses induces a time periodic modulation of the acoustic parameters which may envisage the brain as a time periodic electro-acoustic medium. The temporal varying media framework could open numerous possibilities of wave phenomena to consider. This provides a theoretical background for effects such as the dynamic modification of the propagation phenomena as well as wave generation and amplification. This framework could contribute to the progress of knowledge of the routes involved in the neuronal plasticity and in the consolidation of the memory, as well as the generation of the signals associated to the brain field theory.

## 3. Limitations and particularities of the model

1 – Although in the first part of the work we have calculated de values of the phase velocity, a more realistic model for calculating the group velocity, considering the membrane properties, would be necessary.

2 – Role of other ions $Ca^+$, $K^+$, $Cl^-$ etc. Role of neurotransmitters and other intracell species in the wave propagation phenomena.

3 – The train of pulses are not infinite in time but they are composed by a finite number of pulses. Therefore, broadening of the peaks as well as windowing effects are expected. Therefore, a more realistic scheme considering these peculiarities could match better a bounded time scheme.

4 – In the temporal crystal formulation, the spatial terms are not included. Therefore, the time varying compressibility concerns a bounded area (such as the soma) or a certain discrete part of the transmission line [51]. The inclusion of the spatial modulation could add more properties of wave transmission as it has been reported in many theoretical studies [64,65,66] such as non-reciprocity propagation of the electro-acoustic wave. Non reciprocity means that the AP pulse propagates in the forward direction. In any case, a hypothetical additive effect of counter propagating waves (such as back reflections in axon branch points), could create spatial regions of strong effects of time varying media.

## 4. Acknowledgements

Authors would like to acknowledge to Prof A. J. Hudspeth from Howard Hughes Medical Institute and Laboratory of Sensory Neuroscience, New York, to Dr. E. Estrada from the ETH, Zürich, to Prof. A. Manjavacas from CSIC, Madrid, to J. Meseguer-Sanchez from the ICMOL at the University of Valencia, to Dr. J. Baixeras and Dr. V. Herranz both from the University of Valencia and to Prof. J Sanchez-Dehesa from the Univeritat Politecnica de Valencia, for critically reading the manuscript as well as for their useful and wise comments. We also thank Dr. H. Estrada for bringing to our attention reference (63). Finally, we thank Prof. Zurita-Sanchez from the University of Puebla, Mexico for his help concerning computation aspects of the temporal acoustic model, as well as to A. Ramiro-Soba for his help concerning technical details on potential experiments.

Correspondence and requests for materials should be addressed to F.M. (e-mail: fmese@fis.upv.es) and/or F.R.M. (e-mail: ferraman@fis.upv.es) .

# SUPPORTING INFORMATION.

## S1. Calculation of the neuron stiffness parameter

The sound velocity, $c$, in a fluid is given by the following expression,

$$c = \sqrt{B/\rho}; \; B = 1/\kappa \quad (S1)$$

being $B$, $\kappa$ and $\rho$ the bulk modulus, compressibility, and density values of the fluid medium system respectively.

As the neuron is composed by the neuron fluid and the axon membrane the compressibility parameter associated to the AP, $\kappa$, can be written as

$$1/\kappa = 1/\kappa_M + 1/\kappa_{ion} \quad (S2)$$

Where $\kappa_M$ are the lateral area isothermal compressibility of the axon membrane, and $\kappa_{ion}$ is the compressibility factor associated to the Na$^+$ cation cloud. As reported in the main text we assume the compressibility comes from the electrostatic repulsion forces within the intracell ion cloud. Then $\kappa_{ion}$ can be written as [27]

$$\kappa_{ion} = -\frac{1}{V}\left(\frac{\partial V}{\partial P}\right)_T \quad (S3)$$

where $P$ is the Coulomb repulsion/attraction pressure of ions and $V$ the volume of the unit cell for a given ion distribution at the intracell or extracell fluid.

As discussed in the main body of the text, we will consider two different models both of them based on Coulomb interaction of ions:

A) Neuron deformation is originated from the extracellular Debye layer of Na$^+$ ions attached at the outer side of the neuron membrane [23].
B) Neuron deformation is originated from the electrostatic repulsion [26] within the intracellular layer of Na$^+$ ions at the inner side of the neuron membrane.

In both models we have assumed a total voltage difference between the depolarized and polarized voltage values of the action $V \approx 100 mV$.

### A. Neuron deformation is originated from the extracellular Debye layer of Na+ ions attached at the outer side of the neuron membrane (see Figure 1a).

In this case the ion cloud is located as a Debye layer at the extracellular space. As the Debye layer length $\lambda_d$ is very small of the order of $\lambda_d \approx 1 \; nm$, we assume the Na$^+$ ions are located as a monolayer ordered in a cubic lattice with a nearest neighbor distance (periodicity value), $d$. Therefore, in this periodic arrangement the unit cell volume is $V = d^2 \lambda_d$, Each ion within the ion cloud pushes out the neuron membrane with a electrostatic force $F_C$ [26],

$$F_C = \frac{e^2}{4\pi\epsilon_0\epsilon'\lambda_d^2} \quad (S4)$$

The pressure on the membrane can be written as $P_C = F_C/d^2$. As Na$^+$ ions in the Debye layer are free to move along the surface membrane (z direction), we assume the effective axial pressure, P, is $P = \frac{2}{3}P_C$ [67] (see Figure S1a). So, the electrostatic tensile pressure on the membrane P, will be,

$$P = \frac{2e^2}{12\pi\epsilon_0\epsilon' d^2 \lambda_d^2} \quad (S5)$$

being $d$, the nearest neighbor of the Na$^+$ intracell ions. By replacing (S5) into (S3) we obtain the following expression of the compressibility $\kappa_{ion}^A$

$$\kappa_{ion}^A = 3\pi\epsilon_0\epsilon'\lambda_d^2 d^2/e^2 \quad (S6)$$

being $\epsilon_0$, $\epsilon'$, and $e$ the dielectric constants of the vacuum, the relative dielectric constant of the fluid and the electronic charge of Na$^+$ ion respectively. We have taken $\epsilon' = 80$ [10]. From expressions (S1) we can calculate the phase velocity associated to this model A depicted in Table 1 and Table S1. The compressibility parameter $\kappa_{ion}$ expression (S6) depends exclusively on the inter ionic distance, $d_s$, and not on the neuron diameter. A more realistic model which is out of the frame of this work would consider other parameters as the neuron diameter as well the surface viscosity of the neuron membrane.

From (S5 and S6) we can calculate the ratio between the AP pulse velocity of myelinated neurons over non-myelinated ones.

$$c^B_{myel}/c^B_{unmyel} = \sqrt{\kappa^B_{ion}(unmyel)/\kappa^B_{ion}(myel)} = \left(N^B_{myel}/N^B_{unmyel}\right)^{1/2} \quad (S7)$$

## B. Neuron deformation from the electrostatic repulsion within the intracellular layer of Na⁺ ions at the inner side of the neuron membrane.

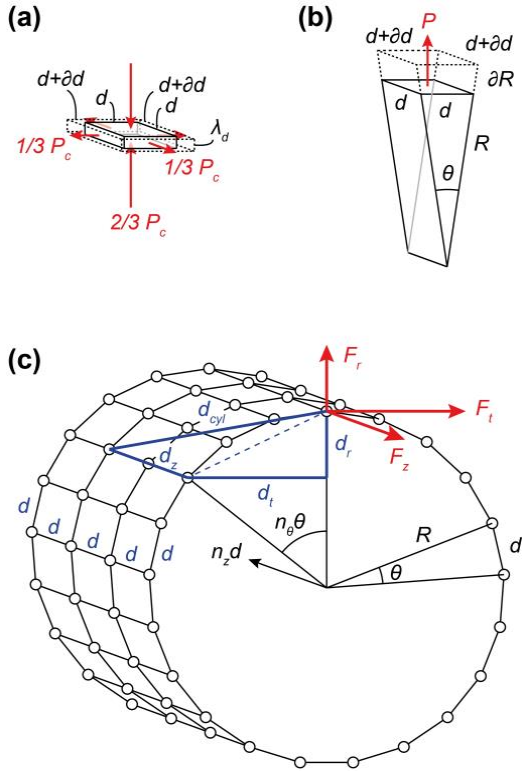

*Figure S1. Schemes of the considered elementary volume for A (panel a) and B (panel b). Panel c depicts intracell $Na^+$ ion cloud arrangement in model B.*

In this model, at variance to the model A, repulsion forces are considered and no Debye layer appears at the inner side of the membrane. Here, we assume the ion cloud is arranged as a curved two-dimensional layer attached at the inner side of the neuron. This permits to consider the radial (out-of-plane) component of the force that produces the swelling of the axon. In contrast with the A model where the attractive force is defined by the Debye layer thickness, here the multiple contribution of surrounding neighbors induces the resultant force. We assume a simple square lattice arrangement of Na⁺ ions with periodicity value $d$ (see Figure S1b,c). We assume the effective area occupied by the each charge is $A = d^2$. The Coulomb repulsion force $F$ on each ion coming from the rest of the ion cloud can be written as

$$F = \frac{e^2}{4\pi\epsilon_0\epsilon' d^2} \cdot s \quad (S8)$$

The factor $s$ has been added to consider the relative interactions among neighbor ions. In particular, $s$ include terms of force projection as well as terms of relative distance (in a scheme of inverse squared proportion).

In the cylindrical distribution of ions (see Figure 1c, the distance between ions ($d_{cyl}$) could be defined by its Euclidian projections such as the radial ($d_r$), transversal ($d_t$) and longitudinal ($d_z$) distances. Those components are related to the cylindrical coordinates by the following expressions (see Figure S1c):

$$d_r = R(1 - cos(n_\theta \theta)) \quad (S9),$$
$$d_t = R sin(n_\theta \theta) \quad (S10)$$
$$d_z = n_z d \quad (S11),$$

where the cylindrical coordinates corresponds to $n_\theta \theta$ (angular), $R$ (radial) and $n_z d$ (axial), being $n_\theta = 1,2,3, ... N_\theta$ ($N_\theta = 2\pi R/d$) and $n_z = 0,1,2, ... N_z$ the index of the angle between charges ($\theta$) and the index of axial distance, respectively being their limits $N_\theta$ and $N_z$, respectively. Notice that $n_\theta = 0$ is not considered because the ions perfectly aligned in the axial direction do not contribute to any radial effect. The angle between two nearest neighbor charges could be considered as $\theta \approx \frac{d}{R}$ because $d \ll R$. Here, both *cosine* and *sine* functions are not simplified (employing Taylor series) because $n_\theta \theta$ could reach large values.

Then, the distance between two arbitrary ions in the cylinder is (see Figure S1c):

$$d_{cyl} = \sqrt{d_r^2 + d_t^2 + d_z^2} = R\sqrt{2(1 - cos(n_\theta \theta)) + n_z^2 \theta^2} = R \cdot d_{cyl}' \quad (S12),$$

$$\text{were } d_{cyl}' = \sqrt{2(1 - cos(n_\theta \theta)) + n_z^2 \theta^2} \quad (S13)$$

is the relative distance between ions with respect to the radius $R$.

Dividing the distance components $d_r$ and $d_t$, and $d_z$, equations (S9-S11), with respect to the total distance $d_{cyl}$, equation (S12), we can obtain the relative projection components (or projection weights) onto the radial ($w_r$), transversal ($w_t$) and axial ($w_a$) directions as follows:

$$w_r = \frac{1 - cos(n_\theta \theta)}{d_{cyl}'}, \quad (S14)$$

$$w_t = \frac{sin(n_\theta \theta)}{d_{cyl}'}, \quad (S15)$$

$$\text{and } w_z = \frac{n_z d}{d_{cyl}'}. \quad (S16)$$

The next expression:

$$w_{d^2} = \frac{\theta^2}{d_{cyl}'^2}, \quad (S17)$$

corresponds to a relative distance (in a scheme of inverse squared proportion), that multiplied by $\frac{1}{d^2}$ permits to obtain the squared inverse distance between ions,

$$\frac{1}{d^2} \cdot \frac{\theta^2}{d_{cyl}'^2} = \frac{1}{d_{cyl}^2}. \quad (S18)$$

By inserting equation (S14)-(S16) into (S17), the relative contributions to the radial ($s_r$), tangential ($s_t$) and axial ($s_z$), could be obtained:

$$s_r = \frac{w_r}{w_{d^2}} = \frac{\theta^2(1 - cos(n_\theta \theta))}{d_{cyl}'^3}, \quad (S19)$$

$$s_t = \frac{w_t}{w_{d^2}} = \frac{\theta^2 sin(n_\theta \theta)}{d_{cyl}'^3}, \quad (S20)$$

$$s_z = \frac{w_z}{w_{d^2}} = \frac{\theta^2 n_z d}{d_{cyl}'^3}. \quad (S21)$$

By inserting $s_r$, $s_t$ and $s_z$ (equations (S19-S21)), into equation (S8), the expression of radial ($Fr$), tangential ($Ft$) and axial force ($Fz$) can be obtained (see figure S1c):

$$F_r = \frac{e^2(1 - cos(n_\theta \theta))}{4\pi\epsilon_0 \epsilon' d_{cyl}^2 d_{cyl}'}, \quad (S22)$$

$$F_t = \frac{e^2 sin(n_\theta \theta)}{4\pi\epsilon_0 \epsilon' d_{cyl}^2 d_{cyl}'}, \quad (S23)$$

$$\text{and } F_z = \frac{e^2 n_z d}{4\pi\epsilon_0 \epsilon' d_{cyl}^2 d_{cyl}'}. \quad (S24)$$

Therefore, for the superposition of the effect of the considered charges, *s* results in:

$$s = \sum_{n_\theta=-N_\theta, n_\theta \neq 0}^{N_\theta} \sum_{n_z=-N_z}^{N_z} \frac{1-\cos(n_\theta\theta)+\sin(n_\theta\theta)+n_z d}{{d_{cyl}'}^3} \theta^2 \quad . \quad (S25)$$

Here, the tangential and axial contribution are cancelled due to the symmetry of the system. Therefore, by substituting $d_{cyl}'$ from equation (S13) into equation (S25) we obtain:

$$s = \sum_{n_\theta=-N_\theta, n_\theta \neq 0}^{N_\theta} \sum_{n_z=-N_z}^{N_z} \frac{1-\cos(n_\theta\theta)}{\left(2(1-\cos(n_\theta\theta))+n_z^2\theta^2\right)^{3/2}} \theta^2 \quad . \quad (S26)$$

Assuming an axon of radius $R \approx 0.5\ \mu m$, the *s* value converges for a large $N_z$ values to $s \approx 6.27$. Considering the electrostatic pressure *P* against the inner side of the neuron can be written as

$$P = \frac{F}{S} = \frac{e^2}{4\pi\epsilon_0\epsilon' d^4} \cdot s, \quad (S27)$$

the effective volume *V* of a single charge as $V = \frac{Rd^2}{2} \approx \frac{d^3}{2\theta}$ (see Figure S1b), and equation (S3), after a bit of algebra, we obtain the following formula for the compressibility factor $\kappa_{ion}^B$:

$$\kappa_{ion}^B = \frac{3\pi\epsilon_0\epsilon' d^4}{e^2 s} \quad (S28)$$

From expressions (S1) we can calculate the phase velocity associated to this model B. The Table S1 provides details of the calculated parameters for both model A and B including their individual acoustic speed. Both models provide phase velocity values of similar value to those appearing in the literature. As described in the core text, $\kappa_{ion}^B \gg \kappa_{ion}^A$ therefore $\kappa_{ion}^A$ has a dominant role in the calculation of the phase velocity for the case of myelinated axons (table 1).

| Cell membrane | Q (C/cm$^2$) | σ (cm$^{-2}$) | d (nm) | $\kappa^A_{ion}$ (Pa)$^{-1}$ | $c^A$ (m/s) | $\kappa^B_{ion}$ (Pa)$^{-1}$ | $c^B$ (m/s) |
|---|---|---|---|---|---|---|---|
| depolarized | 1.0·10$^{-7}$ | 6.3·10$^{11}$ | 12.6 | 2.1·10$^{-5}$ | 6.9 | 1.1·10$^{-3}$ | 0.97 |
| half-depolarized | 5.0·10$^{-8}$ | 3.2·10$^{11}$ | 17.9 | 4.2·10$^{-5}$ | 4.9 | 4.3·10$^{-3}$ | 0.48 |
| polarized | 1.0·10$^{-8}$ | 6.3·10$^{10}$ | 40.0 | 2.1·10$^{-4}$ | 2.2 | 1.1·10$^{-1}$ | 0.10 |

*Table S1. Details of the parameters evaluated for calculating the sound speed where Q (Coulomb/cm$^2$) is the Na$^+$ ion concentration, σ is the Na$^+$ ion layer density, d is the nearest neighbor Na$^+$ ion distance in nanometers. It also shows the compressibility parameters of both models, as well as the total compressibility $\kappa_{ion}$ and the phase velocity, c. We have a assumed a depolarization voltage of 0.1 Volt, a membrane capacitance C= 1.0 µF/cm$^2$ and the relative parameter $s \approx 6.27$ for the B model.*

## S2. The neuron as a time periodic medium. Mathematical development

The sound wave equation based on the continuity equation and the Euler equation can be written as [27]

$$\frac{\partial^2}{\partial z^2} p_R(z,t) - \frac{\partial^2}{\partial t^2}\left[\frac{p_R(z,t)}{(c(z,t))^2}\right] = 0 \quad \text{(S29)}$$

$$c^2 = 1/\rho\kappa = c_0^2 \kappa_0/\kappa \quad \text{(S30)}$$

being $c_0$, $\kappa_0$ the sound velocity and the compressibility of the intracellular fluid at the polarized stage, respectively. We assume the fluid density, $\rho$, is a constant parameter (we can assume the fluid is essentially water).

Equation (S29) has a plane wave solution of the form,

$$p_R(z,t) = p_R(t) e^{ikz} \quad \text{(S31)}$$

Where $k$ is the sound wave vector. After including (S31) into the wave equation (S29) we obtain the following expression

$$k^2 c_0^2 p_R(t) + \frac{\partial^2}{\partial t^2}[p_R(t)\kappa'(t)] = 0, \quad \text{(S32)}$$

$$\kappa'(t) = \kappa(t)/\kappa_0 \quad \text{(S33)}$$

where $\kappa'(t)$ is the compressibility contrast (or relative compressibility) defined as the ratio between the time varying compressibility of the intracell fluid at the action potential stage, $\kappa(t)$ (membrane depolarization) and the compressibility constant $\kappa_0$ when the neuron membrane is polarized.

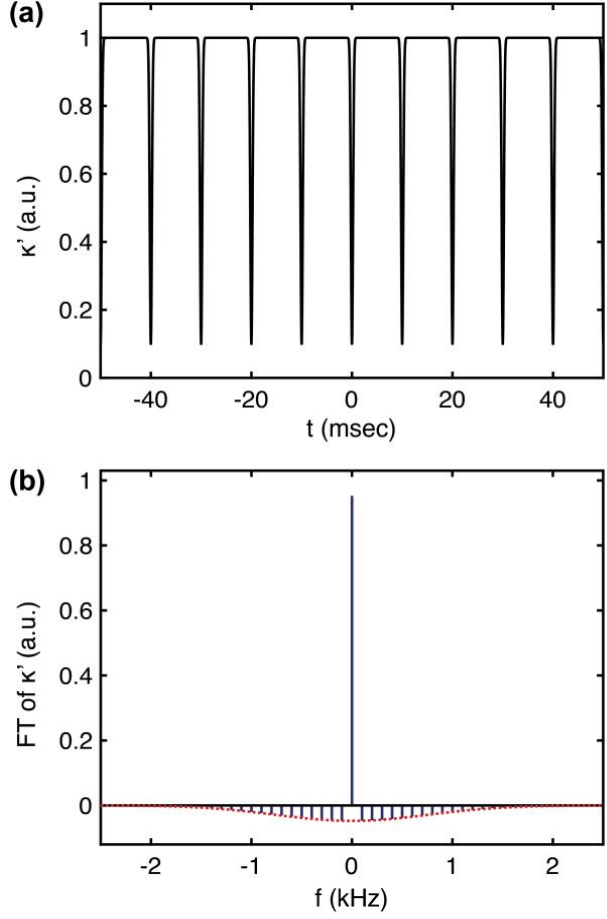

Fig S2. (a) Time domain relative compressibility ($\kappa'$), being their minimum amplitude $\kappa' = 0.10$. (b) Fourier Transform (blue lines) of (a) and its envelope (dotted red curve)

After an algebraic development like that of Temporal Photonic Crystals [18] we get the following canonical equation,

$$\sum_m \left[\sum_{p-m} \alpha_p(\omega) \kappa'_{p-m}(\omega) - k^2 c_0^2 \delta_{pm}\right] q_m(\omega) = 0 \quad \text{(S34)}$$

$$p, m = 0, \pm 1, \pm 2, \pm 3 \ldots$$

$$\alpha_p(\omega) = (\omega - p\Omega)^2$$

being $\delta_{pm}$ the Kronecker delta tensor; $\kappa'_{p-m}(\omega)$ and $q_m(\omega)$ the Fourier components of both the relative compressibility and the sound wave pressure respectively. Expression (S34) corresponds to a set of linear equations which connect all the Fourier components of the membrane pressure amplitude $q_m(\omega)$, which are the eigenvalues of the equation (S32).

Equation (S34) can be regarded as the interaction of an acoustic signal of a frequency $\omega$ propagating in a medium (the neuron) with a relative compressibility parameter, $\kappa'$ periodic in time originated by the periodic train of pulses of frequency $\Omega = 2\pi/T$. Then, we must introduce the Fourier components of $\kappa'$ into equation (S34). It can be modelized in the time domain as the addition of a train of gaussian functions (see Figure S1a) like,

$$\kappa'(t) = 1 + (\kappa'_M - 1) \sum_n \exp\left(-\frac{(t-nT)^2}{2\delta^2}\right) \quad \text{(S35)}$$

$$n = 0, \pm 1, \pm 2, \pm 3 \ldots$$

where $\kappa'_M = \frac{\kappa(unpolariced)}{\kappa(polariced)} = 0.10$ (seeTable 1) , $\delta = FWHM/(2\sqrt{2ln2})$ and $FWHM$ is the full width at half maximum of a single pulse.

In order to get the $\kappa'_{p-m}(\omega)$ appearing in expression (S34) we need to consider the Fourier transform (FT) of $\kappa'(t)$, (see Figure S2(b)).

While a FT of a simple Gaussian function is another Gaussian function, the FT of a train of pulses is a discrete set of values (blue lines of Figure S2b). These Fourier components follow a Gaussian envelope (dotted red curve) and their spectral separation corresponds to the firing rate (assumed to be $1/T = 100\ Hz$). The Gaussian envelope of the FT is inversely proportional to the FWHM of a single pulse: i.e.: narrow pulses in the time domain corresponds to pulses which expands along a large frequency range, $\Delta\omega$. Through all the text we have assume an AP pulse whose FWHM value is $5 \cdot 10^{-4}$sec.